\documentclass[manuscript,screen]{acmart}
\AtBeginDocument{%
  \providecommand\BibTeX{{%
    \normalfont B\kern-0.5em{\scshape i\kern-0.25em b}\kern-0.8em\TeX}}}

\usepackage{wrapfig}

\begin{document}


\title[The Impact of Generative AI Coding Assistants on\\ Developers Who Are Visually Impaired]{The Impact of Generative AI Coding Assistants on\\ Developers Who Are Visually Impaired}

\author{Claudia Flores-Saviaga}
\email{floressaviaga.c@northeastern.edu}
\affiliation{%
  \institution{Northeastern University}
  \country{USA}
}

\author{Benjamin V. Hanrahan}
\email{benhanrahan@microsoft.com}
\affiliation{%
  \institution{Microsoft}
  \country{USA}}

\author{Kashif Imteyaz}
\email{imteyaz.k@northeastern.edu}
\affiliation{%
  \institution{Northeastern University}
  \country{USA}
}

\author{Steven Clarke }
\email{stevencl@microsoft.com}
\affiliation{%
 \institution{Microsoft}
 \country{United Kingdom}}
\author{Saiph Savage}
\email{s.savage@northeastern.edu}
\affiliation{%
  \institution{Northeastern University \& Universidad Nacional Autonoma de Mexico (UNAM)}
  \country{USA \& Mexico}}
\renewcommand{\shortauthors}{Flores-Saviaga, et al.}

\begin{teaserfigure}
\centering
\includegraphics[width=0.8\columnwidth]{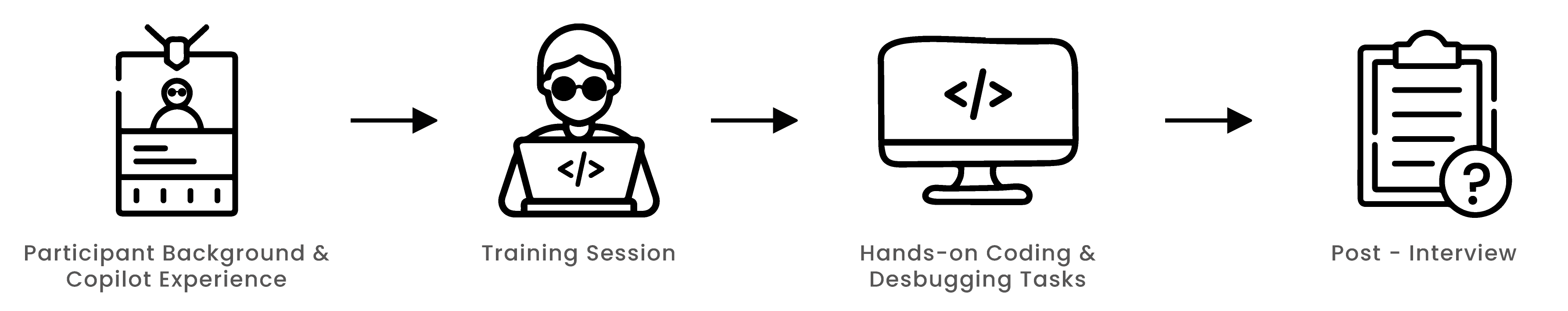}
\caption{Overview of our research to analyze how developers who are visually impaired interact with AI coding assistants.}
\label{fig:setup}
\end{teaserfigure}
\begin{abstract}
The rapid adoption of generative AI in software development has impacted the industry, yet its effects on developers with visual impairments remain largely unexplored. To address this gap, we used an {\it Activity Theory} framework to examine how developers with visual impairments interact with AI coding assistants. For this purpose, we conducted a study where developers who are visually impaired completed a series of programming tasks using a generative AI coding assistant. We uncovered that, while participants found the AI assistant beneficial and reported significant advantages, they also highlighted accessibility challenges. Specifically, the AI coding assistant often exacerbated existing accessibility barriers and introduced new challenges. For example, it overwhelmed users with an excessive number of suggestions, leading developers who are visually impaired to express a desire for ``AI timeouts.'' Additionally, the generative AI coding assistant made it more difficult for developers to switch contexts between the AI-generated content and their own code. Despite these challenges, participants were optimistic about the potential of AI coding assistants to transform the coding experience for developers with visual impairments. Our findings emphasize the need to apply {\it activity-centered design} principles to generative AI assistants, ensuring they better align with user behaviors and address specific accessibility needs. This approach can enable the assistants to provide more intuitive, inclusive, and effective experiences, while also contributing to the broader goal of enhancing accessibility in software development.

\end{abstract}

\maketitle

\section{Introduction}
The integration of generative artificial intelligence (AI) into software development is fundamentally transforming coding practices \cite{liang2024large}. AI coding assistants, such as GitHub Copilot \cite{githubCopilot}, provide functionalities like auto-completion, code generation, and test generation \cite{copilot_vs_cpp_features}, which have the potential to revolutionize how developers work \cite{peng2023impact}. Amidst this surge of new capabilities driven by AI, an important question emerges: How do these generative AI coding assistants impact developers with visual impairments?

The convergence of AI and accessibility in coding presents both tremendous opportunities and intricate challenges \cite{liang2024large, nguyen2024beginning}. On one hand, AI has the potential to make coding more accessible for developers with visual impairments by reducing manual coding tasks and providing intelligent assistance \cite{10.1145/3555609}. On the other hand, if these AI coding assistants are not designed with accessibility in mind, they may inadvertently create new barriers or worsen existing ones \cite{nguyen2024beginning, wang2024investigating}. This risk is especially high if the AI coding assistants depend heavily on visual cues or employ interaction methods incompatible with screen readers \cite{sharif2021understanding, morris2018rich, zong2022rich}. 

In this context, it is important to understand that the relationship between accessibility and usability in AI assistants mirrors longstanding challenges in web accessibility \cite{henry2014role,lazar2004improving}, particularly for visually impaired users \cite{bigham2010vizwiz,kane2008slide}. Similar to rich internet applications \cite{antonelli2019challenges,paahlstorp2007exploring}, AI assistants often introduce dynamic content and complex interactions that traditional accessibility approaches may not fully address \cite{kelly2005implementing,hou2024large}. For instance, Petrie and Kheir \cite{petrie2007relationship} found that existing accessibility guidelines fail to capture many of the usability problems encountered by visually impaired users when interacting with dynamic content. This gap could be especially exacerbated in AI coding assistants \cite{mowar2024tab,oswal2024examining,adnin2024look}, where AI-generated suggestions, real-time code generation, and sophisticated user interfaces can present new hurdles \cite{hou2024large,jalil2023transformative}. Just as screen readers struggle with dynamic web content, AJAX updates, and automatic refreshes \cite{borodin2010more}, developers who are visually impaired may experience similar difficulties when interacting with rapidly changing AI-generated code suggestions and interface elements \cite{li2025prompting}. Consequently, simply following basic accessibility rules might not be enough to make AI coding assistants truly easy to use. We might need to rethink how to make these assistants more accessible for everyone.  The challenge is that we do not fully understand the difficulties and benefits that visually impaired developers experience when using AI coding assistants. Closing this knowledge gap is important—not just for accessibility, but also for making AI-assisted coding tools more user-friendly and effective for all developers, including those with different levels of vision.

Our study addresses this research gap by investigating what unique challenges and opportunities AI-assisted coding tools present for developers with visual impairments. Our research is driven by the following research question:
\begin{itemize}
    \item RQ1: What challenges do AI-assisted coding tools pose for developers who are visually impaired, and what opportunities do they offer to empower and enhance their work?
\end{itemize}

To study this question, we use the Activity Theory framework \cite{kaptelinin2009acting}, which helps analyze how tools influence human activity and identifies \textit{contradictions} that occur when a tool's design does not fit well with users' workflows and needs \cite{baldwin2020activity}. We applied this framework to examine a qualitative study we conducted with 10 visually impaired developers of varying experience levels. These developers used an AI coding assistant, GitHub Copilot, to complete a coding task. During the study, we observed their real-time interactions with the AI assistant, noting both challenges and opportunities. After the task, we conducted interviews to gain deeper insights into their experiences and the difficulties they faced while using the AI coding assistant.

Our findings reveal a complex landscape where AI coding assistants can both improve coding efficiency and introduce new accessibility challenges. Based on our findings, we outline a roadmap for the next generation of accessible AI coding assistants. By analyzing the experiences of visually impaired developers using AI-assisted coding tools, we derive key design insights that can reshape how accessibility is approached in AI-driven software development.

The contributions of this paper are twofold:

\begin{itemize}
    \item We conduct a comprehensive analysis of the accessibility challenges and benefits of AI coding assistants, using real-world insights from developers who are visually impaired.
    \item We propose a set of design recommendations to make AI coding assistants more accessible and inclusive for developers with visual impairments.
\end{itemize}

\section{Related Work}

\subsection{Accessibility and Tools for Developers who are Visually Impaired}
Accessibility in general has been a subject of ongoing research \cite{bigham2017effects,gleason2018crowdsourcing,gleason2017luzdeploy,savage2022global}. Several studies have shed light on the significant challenges developers who are visually impaired face with coding and the techniques these programmers employ to overcome these challenges. Mountapmbeme et al. \cite{mountapmbeme2022addressing} categorized these challenges into five main areas: code navigation, code comprehension, code editing, code debugging, and code skimming. Albusays and Ludi \cite{albusays2016eliciting} found persistent challenges in code navigation, accessing diagrams, debugging, and UI layout. Their study highlighted a  strong preference for text editors over IDEs due to accessibility issues, reduced complexity, and better compatibility with assistive technologies.

Further studies by Mealin and Murphy-Hill \cite{mealin2012exploratory} and Baker et al. \cite{baker2015structjumper} explored specific tools and techniques that developers who are visually impaired use. Mealin and Murphy-Hill found that  many developers who are visually impaired rely on text editors rather than IDEs due to accessibility issues employing unique practices like ``out-of-context editing,'' where blocks of code are copied, edited separately, and pasted back.  This preference for text editors has been corroborated by other researchers \cite{albusays2017interviews, mountapmbeme2022addressing}, who attribute it to specific challenges such as navigating line-by-line, understanding indentation, and managing nested code structures \cite{ludi2015position,utreras2020accessibility}. To address this issue, Baker et al. \cite{baker2015structjumper} created StructJumper, a plugin that generates a hierarchical tree structure of code for easier navigation. Debugging, presents another significant challenge for developers who are visually impaired, largely due to the reliance on visual interfaces in debugging tools, which screen readers struggle to interpret effectively \cite{albusays2016eliciting,potluri2018codetalk}. These challenges have led the development of tools like Wicked Audio Debugger (WAD) \cite{stefik2007wad2} , a tool that provides audio descriptions of programs during execution; CodeTalk \cite{potluri2018codetalk}, a Visual Studio plugin that introduces ``TalkPoints'' for audio-based debugging; and 
CodeWalk, a tool by Potluri et al. \cite{potluri2022codewalk}, which facilitates accessible, remote, synchronous code review and refactoring activities by tethering collaborators' cursors with the host of a Live Share session.  Additionally, Haque et al. \cite{ehtesham2022grid} introduced Grid-Coding, a paradigm designed to improve the accessibility of coding environments by representing code in a structured 2D grid, allowing developers who are visually impaired to navigate and edit code more effectively.

Conversational interfaces have also shown promise \cite{bigham2017deaf},  Ludi et al. \cite{ludi2016exploration}, found that speech-based cues generally provided the best performance for comprehension and navigation tasks.  Phutane et al. \cite{phutane2023speaking} explored the use of voice commands to reduce cognitive load. Meanwhile, Stefik et al. \cite{stefik2011design} created Sodbeans, an IDE that relies on audio cues for debugging, while Smith et al. \cite{smith2003nonvisual} developed a tool to allow developers to navigate the tree structure of files in Eclipse.

While previous research has significantly advanced accessibility for developers who are visually impaired, these findings may not fully apply to the new landscape shaped by generative AI-assisted coding tools such as GitHub Copilot. The introduction of these tools brings new challenges and opportunities in accessible programming, an area that remains largely unexamined. The majority of current studies were conducted before the emergence of generative AI in coding environments, resulting in a critical knowledge gap regarding the impact and potential of these advanced tools for developers who are visually impaired.

\subsection{AI-assisted Coding Environments}
The integration of AI into software development tools has significantly impacted coding practices, with AI-assisted coding assistants like GitHub Copilot showing promise in improving developer productivity. Ziegler et al. \cite{ziegler2022productivity} found that Copilot increases users' feelings of productivity, with almost a third of its proposed code completions being accepted. In a controlled experiment, Peng et al.\cite{peng2023impact} demonstrated that software developers using Github Copilot were able to complete programming tasks significantly faster than those without such assistance. However, Vaithilingam et al. \cite{vaithilingam2022expectation} noted that while most participants preferred using Copilot in daily programming tasks, they often faced difficulties in understanding, editing, and debugging code snippets generated by Copilot, which significantly hindered their task-solving effectiveness. This was confirmed by Dakhel et al \cite{dakhel2023github}, who  noted that the effectiveness of tools like Github Copilot depends on the developer's level of expertise, as less experienced developers often lack the necessary skills to effectively evaluate AI-generated code suggestions.

A recent survey of developers by Liang et al. \cite{liang2024large} revealed that the primary motivations for using AI programming assistants include reducing keystrokes, finishing programming tasks quickly, and recalling syntax. Developers reported that a median of 30.5\% of their code was written with help from tools like Copilot. 

Studies like those by Barke et al.\cite{mozannar2024reading} and Wu et al. \cite{ma2023ai} provide insight into the varying modes of interaction with AI coding assistants. Barke et al. \cite{mozannar2024reading}  identified ``acceleration mode'' and ``exploration mode'' as two broad categories of Copilot use, while Wu et al. \cite{ma2023ai} compared human-human pair programming with human-AI pair programming, highlighting differences in collaboration dynamics.  However, there is still a significant gap in understanding how these AI tools impact developers with accessibility needs.

\subsection{Activity Theory}
\begin{wrapfigure}{l}{0.55\textwidth}
    \centering
    \includegraphics[width=0.55\textwidth]{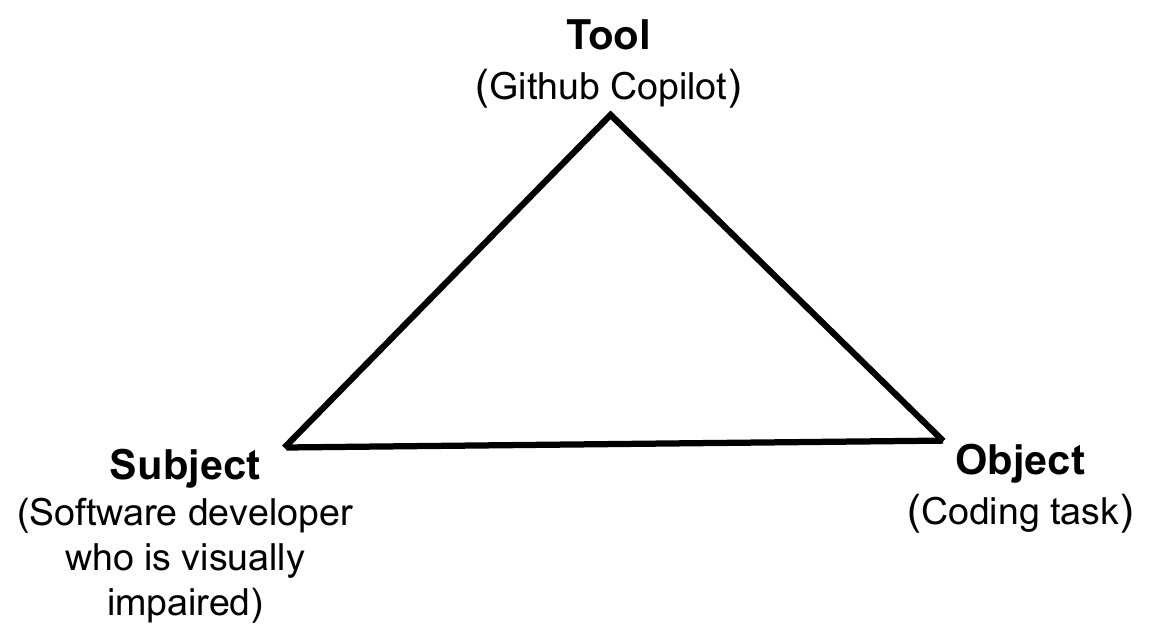}
    \caption{Diagram illustrating the relationship between software developers who are visually impaired (subject), coding tasks (object), and GitHub Copilot (mediating tool) within the Activity Theory framework.}
    \label{fig:activity}
\end{wrapfigure}

Activity Theory is a type of ``conceptual framework'' \cite{ravitch2016reason,rodman1980conceptual,shields2013playbook,soegaard2012encyclopedia}. A conceptual framework is an analytical
tool or structure used to organize and guide
research, projects, or problem-solving efforts \cite{jabareen2009building,miles1994qualitative}. The foundational concept of Activity Theory is the ``activity'', a series of goal directed actions \cite{leont1978activity} that aim to achieve specific objectives \cite{leontiev1981problems}. These actions are mediated by artifacts, the instruments through which individuals interact with their objectives. Actions themselves are performed via routinized operations, in which individuals are not conscious of or focused on these operations. This process is illustrated in Figure \ref{fig:activity}. For example,  developers who are visually impaired (the subjects) may have specific goals (objects) related to completing programming tasks. In this context, an AI programming tool acts as the mediating artifact (tool).

Bødker proposed Activity Theory as a foundational framework for HCI \cite{bodker1989human}, emphasizing its potential to guide the design of interactive systems by focusing on the dynamic interplay between users, their goals, and the tools users employ (interactive systems) \cite{bodker2021through}.
Bødker emphasized that tools should evolve over time to meet users' needs and adapt to the activities they mediate\cite{bodker1989human}, highlighting the importance of designing systems that adjust to changing workflows and contexts to remain effective\cite{bodker2021through}. Within Bødker's definition of Activity Theory for HCI \cite{bodker1989human}, \textit{contradictions} are discrepancies or tensions that arise between the tools, the goals, and the users, that must be addressed \cite{engestrom1987activity}.  These \textit{contradictions} are not obstacles but drivers of change, pointing out areas where tools or processes need to evolve to better meet user needs \cite{engestrom1999activity}. Within this space, Activity Theory also introduces the concept of \textit{misalignments}, which are localized issues or practical mismatches. Unlike \textit{contradictions}, \textit{misalignments} hinder usability and effectiveness for end-users but may not necessitate systemic changes.

In this paper, we use Activity Theory as a framework to examine the design of AI-assisted coding tools for developers who are visually impaired. We chose this approach because Activity Theory has been widely used for decades to study tools and technologies designed for diverse groups, including individuals with disabilities \cite{engestrom1987activity,bodker1989human,kaptelinin2009acting,kaptelinin2012activity,baldwin2020activity,szymczak2023tools,robins2019barrier}. For example, Baldwin et al. \cite{baldwin2020activity} applied Activity Theory to investigate the design of tactile devices and enhanced auditory tools, examining whether these technologies align with the unique workflows of users who have no or low-vision. Similarly, Szymczak \cite{szymczak2023tools} applied Activity Theory to analyze how audio-haptic technologies mediated interactions and supported the goals of individuals who are visually impaired, focusing on how these technologies could assist users in interacting with 2D representations, such as maps and drawings. Tlili et al. \cite{tlili2021game}, also applied Activity Theory to review the use of game-based learning for learners with disabilities and identify inconsistencies in stakeholder involvement, variability in the use of educational technology, and difficulties in standardizing performance measures. Robins \cite{robins2019barrier} applied Activity Theory to analyze and design accessibility in video games, focusing on the interaction between visually impaired players, game goals, and mediating tools like audio-haptic technologies and game mechanics. 

\begin{figure*}[tbp]
\centering
\includegraphics[width=\textwidth, height=1.8\textwidth, keepaspectratio]{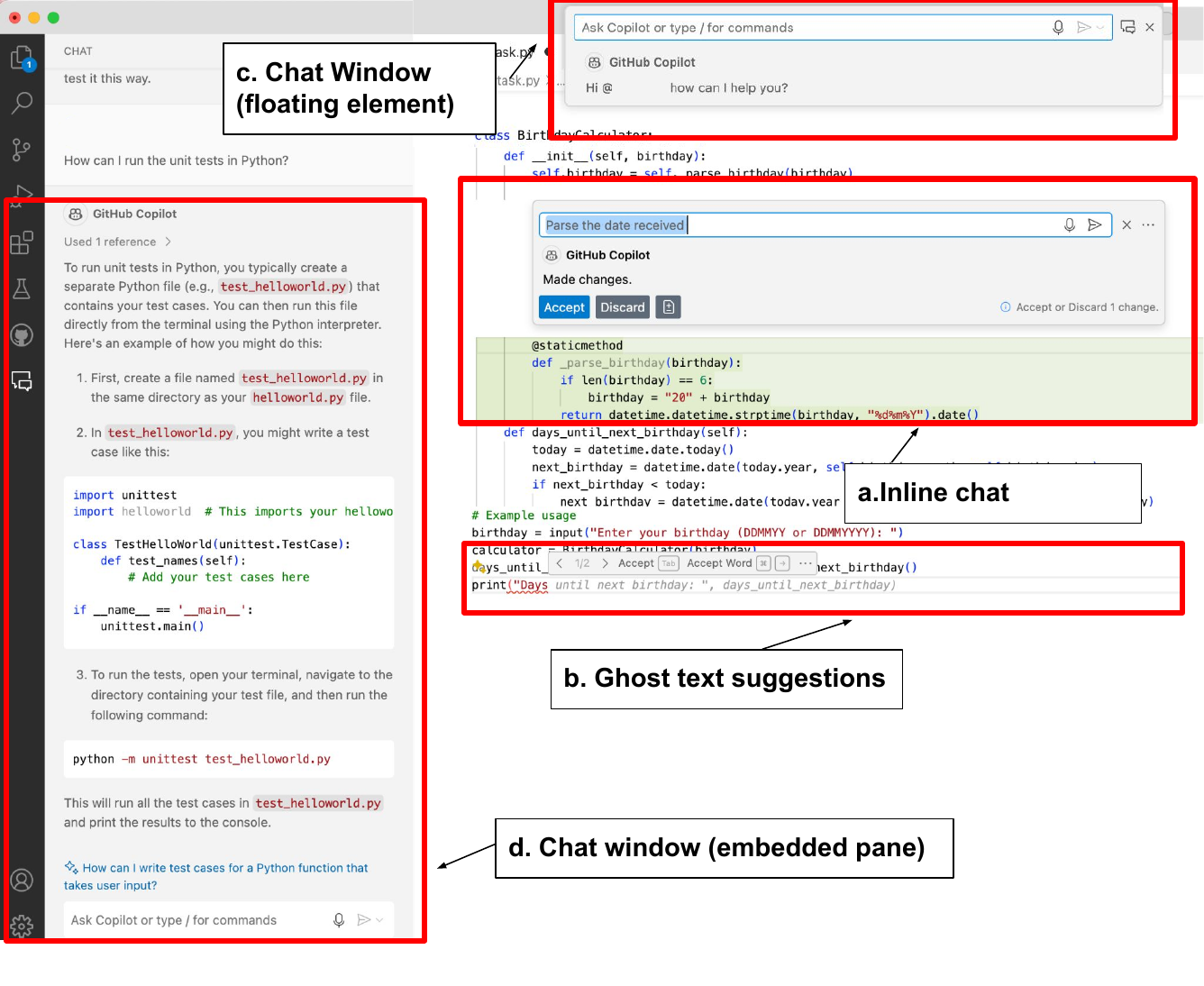}
\caption{GitHub Copilot interaction interfaces. (a) Inline chat window for quick command input and AI engagement. (b) Ghost text suggestions dynamically generated as the user types. (c) Floating chat window for temporary interactions, ideal for quick-access queries. (d) Embedded pane chat window for ongoing, more extensive conversations with AI, allowing for sustained reference and deeper coding assistance.}
\label{fig:copilot}
\vspace{-10pt}
\end{figure*}

\begin{table*}[h]
\scriptsize
\centering
\renewcommand{\arraystretch}{1.2}
\setlength{\tabcolsep}{3.5pt} 
\resizebox{\textwidth}{!}{%
\begin{tabular}{|c|c|c|c|c|c|p{3.5cm}|c|c|}
\hline
\textbf{P\#} & \textbf{Gender} & \textbf{Age} & \textbf{Exp.(yrs)} & \textbf{Vision Status} & \textbf{AI-Coding Tool Experience} & \textbf{Coding Proficiency} & \textbf{Pref. Lang. For Study} & \textbf{Task Completed} \\
\hline
P1  & Male  & 32 & 8  & No vision   & GitHub Copilot & C\#, JavaScript, Python, Ruby & Python & Yes \\
P2  & Male  & 34 & 12 & Low vision  & GitHub Copilot & PHP, C\#, JavaScript & C\# & Yes \\
P3  & Male  & 38 & 9  & Low vision  & CodeWhisperer, Codellama, ChatGPT & C++, Python & Python & Yes \\
P4  & Male  & 43 & 22 & Low vision  & ChatGPT & JavaScript, TypeScript, C\# & C\# & Yes \\
P5  & Male  & 26 & 6  & No vision   & GitHub Copilot, ChatGPT & Python & Python & Yes \\
P6  & Male  & 36 & 17 & Low vision  & GitHub Copilot & PHP, Ruby, Swift, Go, Python, JavaScript, Kotlin & Python & Yes \\
P7  & Male  & 58 & 23 & No vision   & GitHub Copilot & Python, SQL, PowerShell & Python & Yes \\
P8  & Male  & 54 & 32 & No vision   & ChatGPT & C, R, Ruby, TypeScript, Swift, Python, Kotlin & Python & Yes \\
P9  & Male  & 42 & 4  & No vision   & ChatGPT & JavaScript, PHP, Python, HTML, CSS & Python & Yes \\
P10 & Male  & 24 & 2  & No vision   & Claude, ChatGPT, Gemini & C\#, HTML, PHP & C\# & Yes \\
\hline
\end{tabular}%
}
\caption{Participant Demographics, Vision Status, Experience with AI Coding Assistants, and Programming Background.}
\label{tab:participants}
\end{table*}
Building on this foundation, our paper applies Activity Theory to examine how AI-assisted coding tools, particularly GitHub Copilot, mediate and influence the tasks of developers who are visually impaired. Activity Theory provides a particularly useful analysis framework since the introduction of AI coding assistants may operationalize some actions that have not yet become routine for developers. Additionally, we use Activity Theory to guide design recommendations aimed at improving these tools for this population. To achieve this, we analyze both \textit{contradictions}—systemic tensions within the activity system—and \textit{misalignments}—localized issues that disrupt usability. By addressing these challenges, we ensure AI tools meet developers' needs. 

\subsection{\textbf{Copilot}}
The GitHub Copilot interface provides multiple ways for developers to interact with its AI-driven coding assistant. One primary method is through its \textbf{inline chat} (Fig. \ref{fig:copilot}a), where users can type commands or questions directly within the code editor. This allows for seamless, quick interactions without disrupting the coding workflow. 

Additionally, developers can engage with Copilot using the \textbf{chat window}, which can appear either as a \textit{floating element} (Fig. \ref{fig:copilot}c) or as an \textit{embedded pane} within the interface (Fig. \ref{fig:copilot}d). An embedded pane refers to a UI component that is fixed within the main window, allowing for continuous visibility and interaction without obstructing other elements on the screen. The floating element is useful for quick, temporary queries, while the embedded pane is better suited for extended interactions where users may need to frequently reference past exchanges. This flexibility enables developers to choose an interaction style that best fits their workflow and preferences.

Another key interaction method is through \textbf{ghost text suggestions} (Fig. \ref{fig:copilot}b). As the developer types, Copilot continuously analyzes the context and generates relevant code suggestions in real-time. These suggestions appear as semi-transparent text within the editor, allowing developers to seamlessly integrate them into their code. Developers can choose to accept, modify, or ignore these suggestions based on their needs. Beyond real-time suggestions, developers can also request code completions by writing natural language comments that describe the desired functionality. Copilot then processes these descriptions and generates corresponding code snippets, helping developers implement features more efficiently.

It is important to note that GitHub Copilot is not a standalone code editor but an extension designed to work within Visual Studio Code. The chat windows, including both the floating and embedded pane versions, are features introduced by the GitHub Copilot extension and are not part of Visual Studio Code by default. These chat interfaces allow users to interact directly with AI-generated coding suggestions within their coding environment.

GitHub Copilot was selected as the primary AI coding assistant for this study because, at the time of research, it was the most widely used tool of its kind \cite{github_copilot_announcement,stackoverflow_survey}. We chose Visual Studio Code as the development environment due to its popularity and accessibility, as well as its seamless integration with GitHub Copilot \cite{stackoverflow_survey_2023,github_copilot_vscode}. This setup enabled us to explore how these tools influence the coding experience of developers who are visually impaired.

\subsection{Research Gap}
Despite progress in accessibility for developers who are visually impaired \cite{chemnad2024digital,bhagat2024accessibility,mowar2024tab}, generative AI coding assistants like GitHub Copilot introduce new challenges and opportunities that remain under-explored. Most existing research focuses on traditional coding tools \cite{kulkarni2019digital,abou2018artificial}, overlooking complexities unique to AI-assisted coding. Our study examines these accessibility issues to ensure AI assistants like GitHub Copilot are inclusive and support equal access for visually impaired developers.

\section{Methods}
To study the relationship between AI coding assistants and accessibility, we conducted a study with developers who are visually impaired. Our goal was to understand their experiences, challenges, and strategies when using an AI coding assistant, specifically GitHub Copilot. Using Activity Theory as our framework, we analyzed how Copilot (tool) mediates the interaction between visually impaired developers (participants) and their coding tasks (object). Our analysis highlights the contradictions and misalignments that occur when Copilot's design falls short of meeting the unique needs of developers who are visually impaired, while also uncovering new opportunities that emerge in this environment.

\subsection{Participant Recruitment}
We recruited 10 software developers who were visually impaired, including 4 with low vision and 6 with no vision, with experience ranging from 2 to 32 years in the field (see Table \ref{tab:participants}). Participants were sourced through professional networks, accessibility-focused online forums, and organizations supporting  professionals in tech with visual impairments. To participate, developers needed some familiarity with AI coding assistants, though extensive experience with Copilot was not required. All participants identified as male. Given the specialized nature of developers who are visually impaired, our recruitment pool was naturally limited. However, our sample size aligns with prior HCI studies on underrepresented populations \cite{10.1145/3192714.3192821,nowrin2022exploring,rajaselvi2021survey,creed2024voice,flores2022datavoidant}. Despite these constraints, our study provides valuable insights into the accessibility challenges and opportunities for this underrepresented group. Each participant received \$50 USD (or its equivalent in local currency) as compensation for their time.

\subsection{Study setup} 
We conducted our study remotely, ensuring that participants could use their preferred accessible setup. Each session lasted approximately 90 minutes and followed four sequential phases:

\begin{enumerate}
    \item \textbf{Participant Background and Copilot Experience:} 
    We asked participants to describe their professional and technical backgrounds and share their prior experience with AI coding assistants like GitHub Copilot.

    \item \textbf{Training Session:} 
    We provided participants with a training session that introduced GitHub Copilot and its key features. The session also covered how Copilot integrates with accessible development environments and assistive technologies participants were already using. This ensured that all participants had a solid understanding of how to interact with Copilot before starting the coding tasks.

    \item \textbf{Hands-on Coding and Debugging Tasks:} 
    We instructed participants to complete a coding task followed by a debugging task using GitHub Copilot in Visual Studio Code.

    \begin{itemize}
        \item \textbf{Programming Task:} 
        Following prior work \cite{kazemitabaar2023studying}, we asked participants to write a program that calculates the number of days until the user's next birthday. The program required a birthday string input in either \texttt{DDMMYY} or \texttt{DDMMYYYY} format and a class implementation to handle date calculations. We also asked participants to write unit tests for their class. Throughout the task, we encouraged them to use the "think-aloud" protocol, verbalizing their thought process while interacting with Copilot.

        We selected this task because:
        \begin{itemize}
            \item It incorporates common programming concepts such as string manipulation, date calculations, and object-oriented principles, making it relevant to participants' typical coding workflows.
            \item It does not require specialized knowledge of specific programming frameworks, simplifying participant recruitment.
        \end{itemize}

        \item \textbf{Debugging Task:} 
        We then asked participants to engage in a debugging session, where they identified and fixed errors in the code generated by Copilot. During this session, we prompted them to discuss:
        \begin{itemize}
            \item Their usual approach to evaluating code accuracy.
            \item The methods they used to handle encountered issues.
            \item Whether Copilot simplified or complicated their debugging process.
        \end{itemize}
    \end{itemize}

    \item \textbf{Post-Interview:} 
    After completing the programming and debugging tasks, we conducted a semi-structured interview with each participant. We asked them to explain their approach to organizing their code and describe any adjustments they made based on Copilot's suggestions. They also shared the challenges they encountered and the solutions they implemented. Additionally, we encouraged them to reflect on how Copilot influenced the complexity of their tasks, providing examples of when it was particularly helpful or challenging. Our Appendix contains further details on the post-study interview questions.
\end{enumerate}

A diagram illustrating our study setup is shown in Figure \ref{fig:setup}.

\subsection{Data Collection and Analysis}
We recorded and transcribed all interviews, capturing both verbal responses and relevant audio cues from participants' screen readers and other assistive technologies. We analyzed the data using thematic coding, focusing on significant events that participants experienced while using GitHub Copilot. Two of us independently coded the transcripts, identifying emerging themes and patterns. We paid special attention to pivotal moments that either enhanced or hindered participants' coding processes. These critical incidents provided concrete examples of how Copilot mediated accessibility, revealing specific ways in which the tool aligned with or diverged from the non-visual coding strategies used by developers who are visually impaired. To deepen our analysis, we integrated \textit{Activity Theory}, which allowed us to examine the opportunities and contradictions that visually impaired developers encountered when interacting with an AI coding assistant. In particular, Activity Theory helped us explore within our themes:

\begin{enumerate}
    \item \textbf{Activity-Centric Lens:} how developers who are visually impaired 
 (\textit{subjects}) used GitHub Copilot (\textit{tool}) to complete their coding tasks (\textit{object}). By applying Activity Theory, we mapped the observed dynamics within the activity system, identifying key interactions and tensions among its components. This framework helped us understand how Copilot influenced developers' workflows and where misalignments or challenges emerged.
    
\item \textbf{Systemic Contradictions:} challenges that developers who are visually impaired faced, particularly those arising from systemic \textit{contradictions}. For example, some participants struggled with context switching imposed by the AI, while others had difficulty navigating Copilot’s dynamic behaviors, which disrupted the sequential workflows that are critical for developers who are visually impaired. These disruptions required additional navigation adaptations to maintain productivity.
    
\item \textbf{Adaptive Solutions:} possible adaptive solutions to the challenges we identified. By applying the Activity Theory framework, we gained insights into how AI assistants could be better aligned with the specific workflows of  developers who are visually impaired, ultimately fostering more inclusive and efficient coding environments.
\end{enumerate}

By integrating Activity Theory into our thematic analysis, we gained a deeper understanding of how developers who are visually impaired engage with AI coding assistants. This approach allowed us to identify key accessibility themes and design elements of Copilot that either supported or hindered participants' work processes. Our findings contribute to the broader goal of rethinking accessibility paradigms in AI-assisted coding environments.

\section{Findings}

Through our interviews and observations, we identified key themes that highlight the complex relationship between AI coding assistants, accessibility needs, and coding practices. These themes address our research question by revealing the challenges and opportunities that developers who are visually impaired encounter when using AI-assisted coding tools like GitHub Copilot. 

In this section, we present these themes. For each theme, we include illustrative quotes and describe critical incidents from our interviews. These incidents provide insight into the lived experiences of our participants, offering concrete examples of how generative AI coding assistants can both empower and hinder developers who are visually impaired. We frame our findings within the Activity Theory framework to contextualize these challenges as systemic misalignments between users, their tools, and the coding environments they work in. This approach allows us to examine how AI coding assistants mediate the development process and where breakdowns occur, helping us identify opportunities for more accessible and inclusive AI coding assistants.
\subsection{RQ1: Challenges and Opportunities of AI Coding Assistants for Developers Who Are Visually Impaired}
\subsubsection{\textbf{AI and Control}}
Our interviews revealed that AI coding assistants contributed to a dynamic sense of control among developers who are visually impaired. This enhanced control manifested in several positive ways.  Participants described experiencing a shift toward strategic control over the coding process. Rather than manually performing every task, they found that AI assistant allowed them to delegate routine or less engaging tasks, enabling them to focus on high-level decision-making and overall code structure. This ability to offload work gave them greater control in shaping their development process while maintaining oversight of their projects. For instance, P7 highlighted how Copilot eased their workload and gave them more control over their coding process by handling tasks they found tedious, such as generating docstrings—special multiline comments in programming that explain the functionality of a function, class, or module:
\begin{quote}
\textit{``Overall, my experience with Copilot is positive, especially because it helps me with tasks I don't enjoy. I will write the majority of my own code [...], but where it [Copilot] can help me, I'll definitely let it [Copilot] do it. I love having it generate docstrings for me. That's cool, because frankly, I don't like writing documentation. I'd rather code and let it do the heavy lifting for me...''} - P7, Developer with no vision.
\end{quote}

This experience demonstrates how AI empowers developers by taking over tedious tasks they prefer to avoid. By automating routine work, AI enables developers to maintain greater control over their workflow and focus on more complex and engaging coding challenges. Similarly, participant P1, who had prior experience with AI coding assistants, compared coding with AI to piloting an aircraft, emphasizing how these assistants increase their sense of control:

\begin{quote}
\textit{``I sort of compare it [Copilot] to pair programming almost, where [...] I'm sort of driving, but the other person is doing all the sort of drudgery work. And what I'm really mostly doing now is almost like what a pilot does when they're flying a plane. They're [...] not flying the plane physically for most of it. Most of it, the plane's flying itself, but the pilot does have to keep an eye on. Are we still going in the right direction? What's that big thing coming towards us really fast? Should we avoid that?''} - P1, Developer with no vision.
\end{quote}

From the perspective of Activity Theory, these findings illustrate the transformative role of AI tools as mediators in the coding process. In this context, Copilot functions as an intermediary that actively shapes a developer’s control over their work, allowing developers to reallocate their focus toward higher-level strategic decisions, as demonstrated by P7 and P1. 

This shift in control also aligns with the concept of \textit{supervisory control} \cite{chignell2023evolution,sheridan1992telerobotics}, in which humans oversee and direct automated systems rather than executing tasks manually. In this case, the developers who are visually impaired guide the AI coding assistants by monitoring their outputs, making corrections when necessary, and ensuring alignment with their broader coding goals. Just as a pilot monitors and manages an aircraft’s automated systems while remaining responsible for high-level navigation and safety decisions, developers can delegate routine implementation details to Copilot while maintaining oversight and control over the overall project direction. This dynamic shifts the developer’s role from manually writing every line of code to curating, refining, and strategically guiding the AI’s output. 

In this new paradigm, control manifests as the ability to guide the overall direction of the code, make critical decisions, and intervene when necessary. While the developer retains ultimate authority over the coding process, the nature of that control shifts to a more abstract and strategic level. Rather than focusing solely on writing individual lines of code, developers can now oversee and direct AI-generated suggestions to ensure they align with the broader project architecture and goals. 

However, this shift in control also requires developers to develop new skills, such as crafting effective prompts to generate useful AI outputs, quickly assessing the quality and relevance of the AI-generated code, and maintaining a high-level understanding of the overall software design. As highlighted by P1:

\begin{quote}
\textit{``...when I was checking that initial version of the day counting function, I wasn't sure if it [Copilot] realized that it wasn't always going to be the same year, and it [Copilot] didn't realize that in this case. But I've seen it make that kind of mistake before [...] You just need to be careful with your prompts [communication with the AI assistant].''} - P1, Developer with no vision.
\end{quote}


This shift in AI-assisted coding, where developers focus on strategic oversight rather than manually completing every coding task, marks an evolution in how developers who are visually impaired maintain control over their coding workflows \cite{aljarallah2024systematic}. The challenge lies in designing AI coding environments that empower developers with this level of control while ensuring accessibility.

To achieve this, designers of AI coding environments must recognize that the needs of developers who are visually impaired often differ from those of sighted developers \cite{zhang2023understanding,ehtesham2022grid,aljarallah2024systematic}. While sighted developers frequently rely on visual cues and dynamic, exploratory interfaces, visually impaired developers usually require predictable, structured interactions with clear, interpretable feedback \cite{mountapmbeme2022addressing}. This requirement is not merely a preference but a necessity that allows them to effectively understand, navigate, and maintain control over their workflow \cite{mountapmbeme2022accessible,khalajzadeh2024accessibility}. Now, a key consideration in designing AI coding environments for visually impaired developers is enabling them to anticipate the AI’s actions and seamlessly integrate its assistance into their coding tasks \cite{moured2024chart4blind}. However, generative AI inherently introduces a degree of unpredictability, making this integration complex. As a result, developing AI-driven coding environments that balance strategic oversight with accessibility is not a trivial task and requires careful, thoughtful design.

\subsubsection{\textbf{Context Switching Difficulties in AI-assisted Interfaces}}
Although Copilot provided a sense of control and empowerment, it also introduced new challenges related to context switching. Our findings highlight that the dynamic nature of AI-generated suggestions and the frequent need to shift focus between writing code and reviewing AI outputs created disruptions. For developers who are visually impaired, this shift interfered with the structured navigation strategies they had developed for traditional coding environments \cite{aljarallah2024systematic}, making it harder to maintain workflow continuity. 

The AI assistant frequently triggered unexpected view changes, forcing developers to shift their focus to different tasks or sections of the coding environment. For example, as developers interacted with Copilot's suggestions, the system would automatically switch their view to the newly generated code. This disrupted their workflow, especially because they were not always notified about these sudden shifts. As a result, developers struggled to maintain context, making it even more challenging to navigate and integrate AI-generated code seamlessly.

The challenges of context switching in AI-assisted coding environments pose a significant accessibility barrier for developers who are visually impaired. These difficulties highlight the need for AI coding assistants to be designed with greater attention to the needs of screen reader users, who rely on sequential navigation \cite{hegde2023user, zong2022rich}.

This challenge became evident with P9, who encountered difficulties after typing a question in the embedded chat window. When Copilot generated a response, he struggled to locate and navigate to the text where the answer was displayed, disrupting his workflow and adding unnecessary cognitive load:

\begin{quote}
\textit{``Okay, so do I need, I'm just, I'm trying to understand if we need to press the up and down [the participant repeatedly pressed keys to locate the AI assistant's response]. Seems like it still, the focus is still on my question that I typed [and not on the AI's response]...''} - P9, Developer with no vision.
\end{quote}

P1 emphasized that blind users depend on maintaining a stable and consistent context to work efficiently, largely due to how screen readers are designed. However, the frequent and abrupt context switching introduced by AI coding assistants disrupts this continuity, making it difficult for developers who are visually impaired to stay oriented and manage their coding tasks effectively:

\begin{quote}
\textit{``A blind person really only has one thing they can see at any given time. It would be that one window, one line of text, one line of whatever it is [...] because screen readers, that's just how they work. You can only see one thing at a time. They work sequentially and not parallel.''} - P1, Developer with no vision.
\end{quote}

P5 explained that the context-switching issue worsened due to the many keystrokes required to navigate between different contexts, such as reaching the Copilot chat window after writing code. The need for multiple keystrokes or clicks for minor transitions caused him to lose track of his original task, leading to frustration. This additional effort further amplified the challenge, making it harder to maintain workflow and stay focused on coding tasks:

\begin{quote}
\textit{``...when I'm coding, I have to memorize so much... It's a lot of, I'm going to reference this file again because I can't 100\% remember what I said...I was going to do. And you know when I'm going to Copilot, when I'm pressing F6 a couple of times and I'm pressing tab a couple of times, that creates a little frustration almost, because I want to get back to the task [The participant used multiple key presses to navigate to the Copilot chat window and then return to their code]. And that frustration makes it hard to remember what I was thinking about, to begin with. And then...I forget the context...now I have to go back and check again.''} - P5, Developer with no vision.
\end{quote}


The challenges of context switching in AI-assisted coding environments were further exacerbated by unexpected AI responses that caused sudden and confusing shifts in context. P5, a developer with no vision, described a particularly frustrating experience where merely navigating through the coding interface unintentionally triggered AI actions, making navigation even more difficult:

\begin{quote}
\textit{``Did I accidentally just accept [accept the AI's suggestion] I don't, I think I hit tab. See, this is part of the problem too. It's so easy to accidentally accept a suggestion because when you move the cursor, Copilot immediately is like, oh, I have a suggestion for this and I'm just going to suggest it to you. But you know, again, the way a lot of us [developers who are visually impaired] navigate the UI is usually with the tab key. And so it's so easy to just accidentally insert a suggestion''} - P5, Developer with no vision.
\end{quote}

Similarly, P2 highlighted a significant challenge for users who rely on screen magnification: the AI assistant sometimes displayed information in areas of the screen that were out of view. Because these users were zoomed into a specific section of the screen, they often remained unaware when the AI assistant provided suggestions, as the content appeared outside their visible area:

\begin{quote}
\textit{``...If I, for example, zoom in like this [the participant zoomed into a specific screen area], I often don't get any information if there's some programming error or something going on in this area [the participant pointed to another part of the screen]. When I look at this [the zoomed-in area] and I just remember every time I have to go this way [The participant pointed to the unseen screen area where Copilot placed suggestions.] Sometimes I don't see that [AI suggestion] if I zoom in. So for people who are using Zoom, it will be better that it [AI assistant] will show up in the middle of the program because I don't see it [AI assistant] when it's in this area [zoomed-in area of the screen]. If there's something I have to install, for example, to run anything.''} - P2, Developer with low vision.
\end{quote}

These participant experiences align with and extend prior research on context-switching challenges for developers who are visually impaired. Our findings build upon the work of Albusays et al. \cite{albusays2017interviews}, who conducted interviews and observations with blind software developers to examine code navigation difficulties in traditional IDEs. Their study identified the challenges blind developers face when moving between different sections of a program. Our research expands on these insights by showing how AI-assisted coding environments amplify these difficulties by introducing additional interface elements and new interaction modes, further complicating navigation and workflow continuity. While Albusays et al. \cite{albusays2017interviews} focused on traditional IDEs, our study reveals that AI-assisted coding environments introduce an additional layer of complexity by requiring developers to frequently switch contexts within the same application. This added complexity exacerbates the already challenging task of code navigation for developers who are visually impaired. The dynamic nature of AI-generated suggestions and the frequent need to shift focus between manually written code and AI-generated content disrupt the mental models and navigation strategies that blind developers have developed.

From an Activity Theory perspective, the challenges of context switching reveal \textit{contradictions} in the interaction between the user and the AI coding assistant. These \textit{contradictions} likely arise because AI coding assistants are primarily designed for visual interaction, which conflicts with the sequential, nonvisual navigation methods used by screen reader users. However, several additional factors contribute to this misalignment:

\begin{itemize}
    \item The need for multiple keystrokes to move between different contexts within the AI-assisted environment.
    \item Unexpected AI behaviors that interfere with established navigation patterns of developers who are visually impaired.
    \item Information being displayed in inaccessible areas due to screen magnification.
\end{itemize}

This mismatch between the tool's design and the needs of developers who are visually impaired disrupts their ability to stay focused and maintain a smooth workflow. 

\subsubsection{\textbf{AI Timeouts}}
While AI coding assistants gave developers greater control over their code and reduced the need for manually completing every task, they also introduced cognitive challenges. Participants specifically expressed a need for moments of disconnection from the AI. The constant stream of AI-generated suggestions—especially the frequent appearance of \textit{ghost text}—created a trade-off between maintaining control and mental focus. This tension led to a key theme in our study: participants' desire for what we call \textit{"AI timeouts"}—periods of uninterrupted coding without AI intervention. For instance, participant P1, a braille display user, stressed the importance of having control over when AI suggestions appear. They found that frequent AI interventions disrupted their thought process, stating:

\begin{quote}
\textit{``What I usually do when this happens [when ghost text is presented] is I'll turn speech off for a bit so I can think....''} - P1, Developer with no vision.
\end{quote}

Participant P1's strategy of turning off speech to think highlights the issue of \textit{information overload}, a challenge explored by Ahmed et al. \cite{ahmed2012read} in their research on non-visual interfaces. Ahmed et al. explain that screen-reader users often struggle to determine the relevance or importance of content without first listening to at least some of it. This necessity frequently results in cognitive overload for users who are visually impaired.

In AI-assisted coding environments, this issue becomes even more pronounced. P1's experience illustrates how the continuous stream of AI-generated suggestions, conveyed through speech output, can overwhelm cognitive capacity. By choosing to ``turn speech off for a bit,'' P1 effectively creates a temporary barrier—an \textbf{AI timeout}—allowing them to pause, process information, and make decisions without the constant influx of new suggestions. This aligns with findings by Bigham et al. \cite{bigham2011design}, who suggest that slower-paced interactions can enhance usability for visually impaired users.

The concept of AI timeouts extends beyond merely managing ghost text suggestions. Some participants expressed a need for extended periods of uninterrupted coding without any AI input. For instance, P7 emphasized the importance of having the option to engage in uninterrupted coding sessions, stating:

\begin{quote}
\textit{``Well, hey, can we get a mode that says, hey, I just want to get all my code done. I just want to write code. I don't care right now. I don't care how buggy it is right now. I really don't care. But I want to stay in the IDE because when I'm done, I'm going to use a shortcut and I'm going to go through, I'm going to find all my errors and I'll fix them, but I don't want it [AI assistant] to get in the way [...] But let me just get my thoughts out. ''} - P7, Developer with no vision.
\end{quote}

This experience highlights the potential for cognitive dissonance when AI assistance operates at a faster pace than the developer's thought process. Although the AI's suggestion may have been useful, it momentarily disrupted P7's train of thought, creating a mismatch between the AI’s proactive assistance and the participant’s cognitive workflow. While P7’s experience underscores the need for AI timeouts, it is important to note that not all participants shared this sentiment. P1’s reflection, for example, illustrates a different perspective—one where AI acts as a catalyst for learning and expanding problem-solving approaches:

\begin{quote}
\textit{``I think it [AI assistant] was a couple steps ahead of me now and again because I was sort of just expecting, okay, I need to get that date [date requested in their coding task], and then I'm going to need to somehow get a year later out of it. And I didn't actually really consider that you'd have to sort of parse that date and convert it into an actual date first. [...] So when it [AI assistant] made a constructor and just transformed it into month, day, year, I'm like, okay, well, that's not immediately what I had in mind, but on second thought, that does make a lot of sense. [...] I probably would have done that as a next step myself, but it just sort of preempted me on there.''} - P1, Developer with no vision.
\end{quote}
 
This experience aligns with Vygotsky's concept of the Zone of Proximal Development \cite{vygotsky2012thought}, which suggests that learning is most effective when guidance bridges the gap between what a learner can do independently and what they can achieve with support. In this case, the AI guided P1 toward a more advanced solution. However, maintaining a balance between beneficial cognitive challenges and overwhelming disruptions is crucial. While this instance demonstrated a positive outcome, it highlights the need for AI systems that can dynamically adjust their level of intervention based on the user's expertise and cognitive load. By incorporating adaptive assistance, AI coding assistants can create a more effective and enriching coding experience—one where developers are both supported and challenged in a way that enhances their problem-solving abilities without causing excessive cognitive overload.

Conversely, P4's response to how AI suggestions influenced their planning highlights two key insights. First, AI assistants can encourage developers to adopt best coding practices earlier in the development process. Second, AI timeouts could play a role in giving developers the necessary time to reflect on different aspects of their program before implementing AI-generated suggestions:

\begin{quote}
\textit{``...[I would] not have remembered to put those guard clauses in until I got, like, my unit testing hat on [the participant was pointing to AI-generated guard clauses, a best practice for handling edge cases]. I've seen some videos of like, I think Uncle Bob, Bob Martin did some unit testing videos where he literally had a hat that he would flip around and it was two hats put together and he'd be like, coder, unit test, or coder and those kind of things of like, oh, I wasn't there yet [was not thinking about guard clauses yet], but if you [AI assistant] want to help me get there early, that's fine. That's great.''} - P4, Developer with low vision.
\end{quote}

This reflection highlights how AI can blur the traditional boundaries between different coding phases, potentially reducing the need for certain AI timeouts. In this case, the AI’s proactive suggestions for guard clauses eliminated the need for a ``timeout'' between the coding and testing phases by integrating best practices early in the development process. However, this scenario also emphasizes the importance of  ``flexible AI timeouts''. While P4 valued the early inclusion of guard clauses, other developers might prefer to maintain distinct mental phases in their coding workflow. In this context, AI timeouts do not necessarily mean disengaging from AI assistance entirely but rather enabling developers to control when and how AI interventions occur. By allowing developers to schedule AI timeouts or receive advanced suggestions at specific points in their workflow, AI-assisted coding tools can accommodate different coding styles and personal preferences. This flexibility could be particularly beneficial for developers who are visually impaired, as they may have structured routines or mental models for managing distinct phases of the coding process \cite{latoza2006maintaining,johnson1989mental}. Providing adaptive AI interaction could enhance productivity without disrupting established workflows.

P3's experience further illustrates the need for a balanced approach to AI timeouts, ensuring that AI assistance is helpful without becoming disruptive:
\begin{quote}
\textit{``Honestly, I think it's [AI assistant] super helpful because like, I'm going to be honest, like you were saying before, I always forget datetime manipulation. [...] So having the ability to just say what I want, have it [AI assistant] understand the context of the programming language I'm in and help me kind of get there quicker without having to kind of start searching around the web, figuring out the reading documentation again for the datetime methods. Like I have to all the time.''} - P3, developer with low vision.
\end{quote}

As the quote illustrates, developers may choose to keep the AI when it provides helpful support, such as recalling correct syntax or assisting with complex code manipulations. However, for tasks requiring deep problem-solving or creative thinking, they may prefer to use AI timeouts to independently work through solutions.

The concept of ``AI timeouts'' reflects \textit{contradictions} within the activity system, arising from misalignments between the \textit{subject} (developers), the \textit{object} (coding tasks), and the \textit{tool} (Copilot). While Copilot serves as a mediating artifact designed to facilitate coding, its dynamic and sometimes intrusive interventions create tensions that can disrupt developers' workflows. This contradiction likely arises because the tool's design assumes that continuous AI assistance improves productivity, whereas developers might actually need uninterrupted periods to process and structure information. These tensions highlight the necessity of adaptable AI systems that calibrate their level of intervention based on the user's needs. 

\subsubsection{\textbf{AI-Assisted Coding Efficiency}}
While our study highlighted challenges like context switching, it also revealed key benefits of AI coding assistants, particularly in enhancing coding efficiency for developers who are visually impaired. This efficiency improvement stemmed from two main factors:

\begin{enumerate}
    \item \textbf{Copilot’s Proactive Code Generation}, which anticipated and generated relevant code, reducing manual effort.
    \item \textbf{Copilot as an Accessible, Always-Available Coding Partner}, providing instant non-judgmental support.
\end{enumerate}

\textbf{Proactive Code Generation}. Participants highlighted how the AI assistant boosted their productivity by anticipating their needs—proactively generating code that resolved existing issues or introduced features they had not initially considered. For instance, P7 described a moment when the AI assistant demonstrated foresight by automatically parsing and transforming a date into a usable format. Upon reflection, P7 realized that this step aligned perfectly with their next intended action, streamlining their workflow and reducing the need for manual adjustments:

\begin{quote}
\textit{``It [AI assistant] was a couple steps ahead of me. [The participant's coding task involved calculating the number of days until a birthday provided by an end-user.]  I was initially focused on getting the date and calculating a year later, but I didn't actually really consider that you'd have to sort of parse that date [date given by the end-user] and convert it into an actual date first [converting to an actual date was something the generative AI did for them]. That makes perfect sense in hindsight [...] So when it [AI assistant] made it a constructor and just transformed it into month, day, year, I'm like, okay, well, that's not immediately what I had in mind, but on second thought, that does make a lot of sense. [...] I probably would have done that as a next step myself, but it just sort of preempted me on there.''} - P7, Developer with no vision.
\end{quote}

Similarly, P1 emphasized their appreciation for the AI's ability to proactively generate code, particularly in how it anticipated their next steps and aligned with their intentions as a developer:

\begin{quote}
\textit{``I'm gonna try something else. I'm actually very curious [...] I'll auto modify that [The participant used a Copilot command to 'auto-modify' their code, allowing the AI to proactively edit and expand their initial work]. That sounds good [The participant was reviewing the code generated by the AI'']. There are days until your next birthday. Yeah, format days until next birthday [The participant continued reviewing the code generated by the AI.] That [code generated by the AI] is scarily good. Actually that [AI generated code] is exactly what I was going to do.''} - P1, Developer with no vision.
\end{quote}

Prior research has documented the challenges that developers with visual impairments face when navigating and understanding code structures in traditional development environments \cite{albusays2016eliciting,potluri2018codetalk,baker2015structjumper}. Given these challenges, we argue that AI’s proactive code generation holds potential for this population. Developers who are visually impaired often need to traverse multiple documentation pages or code files to construct appropriate solutions—a process that is not only time-consuming but also more prone to errors when relying on screen readers \cite{albusays2017interviews,borodin2010more}. In this context, AI’s ability to predict coding needs and generate relevant code snippets is especially valuable, as it can reduce cognitive load and enhance efficiency for developers who are visually impaired.

{\textbf{AI as an Accessible, Always-Available Coding Partner.}} Some participants described the AI assistant as a supportive 'buddy' that guided them through their coding tasks, providing assistance and enhancing their productivity. They appreciated having an AI-powered companion they could rely on for coding assistance, providing quick answers and relevant code snippets when needed. For example, Participant P3 likened the AI assistant to a knowledgeable colleague who is readily available to offer solutions and assist with coding challenges:

\begin{quote}
\textit{ ``[with the AI coding assistant] you have that kind of buddy to kind of ask real quick, you know, what was that? You know, what was the daytime method that I needed to use to convert that particular string into the right thing? [The participant refers to asking the AI about functions and methods needed for the date-related coding task]''} - P3, Developer with low vision.
\end{quote}

All participants agreed that the AI assistant was beneficial in helping them complete their software development tasks. However, the degree of reliance on the assistant varied among individuals. Some, like P6, depended on it extensively, using it to complete nearly their entire assigned workload: 

\begin{quote}
\textit{``I just copy and pasted the whole task [instructions for a coding task assigned in our study] into chat [Copilot's chat interface], and it [AI assistant] gave me the whole implementation that we had to do. Very, very minor tweaks left to do. I mean, the date formats worked out, right? [The participant was assigned a coding task involving date formats.] Essentially we just added comments and tinkered with the code [...] But otherwise, it [AI assistant] just did everything for us.''} - P6, Developer with low vision.
\end{quote}

Even participants with limited experience using AI assistants, such as P9, acknowledged its potential to accelerate coding tasks. Their experience in our study sparked greater interest in exploring the technology further:

\begin{quote}
\textit{``I would love to try and explore more [about AI coding assistants]. But, it [the AI coding assistant] has definitely a potential [...] just doing quick work, especially writing quick functions and doing also all those tasks [...] It's [the AI assistant is] bringing a lot of time saving.''} - P9, Developer with no vision.
\end{quote} 

For developers who are visually impaired, this aspect of AI assistance could be particularly valuable, as it offers constant, non-judgmental support without requiring face-to-face interaction. This can be especially beneficial in workplace environments where they may feel hesitant to frequently ask colleagues for help \cite{albusays2017interviews}. Generative AI assistants have the potential to make software development more accessible and efficient, enabling developers to work more independently without relying on coworkers for assistance.

From the perspective of Activity Theory, Copilot serves as a mediating artifact that reshapes the coding activity by shifting its role from a passive tool to an active collaborator. Within the framework of Activity Theory, the interaction between the \textit{subject} (developer), \textit{object} (coding task), and \textit{tool} (AI assistant) is no longer a linear process but a dynamic, adaptive exchange. Instead of developers following a rigid, step-by-step workflow, Copilot enables a more fluid interaction where AI-generated suggestions proactively shape the problem-solving process. Acting as a readily available and non-judgmental coding companion, Copilot can function as a ``buddy'' that developers can turn to at any moment to clarify uncertainties, propose solutions, and reduce cognitive load.  By integrating AI as an active collaborator within the activity system, Copilot can help developers navigate complex programming tasks with greater confidence and independence.

\section{Discussion}
Our findings show that AI coding assistants provide powerful capabilities, new opportunities, and greater control for developers who are visually impaired, but they also introduce new accessibility challenges. Issues such as context switching and managing AI-generated suggestions highlight the need for a new approach to accessibility in AI coding environments. Building on Nielsen’s argument that generative AI has introduced a new paradigm for HCI \cite{nielsen2023ai} and the design principles for generative AI tools outlined by Weisz et al. \cite{weisz2024design}, we propose design recommendations to ensure AI coding assistants are accessible and beneficial for all developers.

\subsection{AI Control and Context-Switching Management in AI Coding Assistants}
Our study highlights the mixed experiences of developers who are visually impaired when using AI coding assistants. While these assistants empower developers by allowing them to focus on higher-level strategic thinking, they also introduce new challenges. Participants, such as P5 and P9, reported difficulties in navigating between their static code and the dynamic windows used for AI interactions. This constant context switching led to frustration, disrupting their workflow and breaking their cognitive focus.

Previous research highlights that the visually oriented nature of Integrated Development Environments (IDEs) presents challenges for developers who are visually impaired, often leading to a loss of control \cite{albusays2017interviews,szpiro2016people,cha2024understanding}. P1 also noted that screen readers require users to read code sequentially, making context switching particularly difficult \cite{albusays2017interviews,mealin2012exploratory,albusays2016eliciting,albusays2020role,smith2003nonvisual}. Several tools, such as StructJumper and CodeTalk, have been designed to aid navigation in traditional static coding environments \cite{baker2015structjumper,potluri2018codetalk,potluri2022codewalk,armaly2018audiohighlight}. However, these tools primarily focus on static code elements. In contrast, AI-assisted coding environments introduce new challenges by incorporating dynamic code elements \cite{peng2023impact}. Our findings reveal that developers who are visually impaired must now manage the additional complexity of interacting with AI-generated suggestions, making context switching even more difficult. Navigating between static code and AI-generated content requires frequent shifts in focus across different interface elements \cite{mozannar2024reading,peng2023impact}. Developers must assess their own code, review AI-generated suggestions, and interact with various UI windows required for AI engagement \cite{wong2023natural}. This continuous back-and-forth process can increase cognitive load, forcing developers to constantly evaluate AI outputs while maintaining an understanding of their overall code structure \cite{russo2024navigating}. As a result, the interplay between AI assistance and manual coding can fragment attention, disrupt workflow continuity, and impact productivity, code quality, and trust in the AI-generated content \cite{peng2023impact,liang2024large,wang2024investigating}.

Our findings highlight the urgent need for accessible tools specifically designed for dynamic coding environments. These tools should aim to streamline workflows, minimize AI disruptions, and provide better support for developers who are visually impaired. By addressing the challenges of context switching and maintaining control, such tools can help developers navigate AI-assisted coding environments more effectively and improve overall accessibility.

\subsubsection{Addressing the Research Gap}
Our research revealed both new accessibility opportunities and challenges in AI coding assistants for developers who are visually impaired. On one hand, AI coding assistants empowered developers by shifting their role toward supervisory control over the code. However, these benefits were accompanied by significant challenges, especially related to context switching, which disrupted their workflow and made navigation more complex.

While previous studies \cite{albusays2017interviews,szpiro2016people,cha2024understanding} have identified accessibility barriers in traditional IDEs—such as visually-oriented interfaces and linear screen reader navigation for static code \cite{mealin2012exploratory,albusays2016eliciting,albusays2020role,smith2003nonvisual}—our findings reveal that AI-driven coding environments exacerbate these issues. The dynamic nature of AI-assisted coding introduces additional complexities that amplify accessibility barriers. Existing tools like StructJumper and CodeTalk \cite{baker2015structjumper,potluri2018codetalk,potluri2022codewalk,armaly2018audiohighlight}, designed for static code, may not fully support AI-driven workflows. Furthermore, prior research on AI-assisted coding has examined cognitive strain and productivity for general developers \cite{peng2023impact,mozannar2024reading,wong2023natural,liang2024large}. But, our work introduces an accessibility perspective. Frequent context switching and fragmented workflows present significant challenges for developers who are visually impaired, emphasizing the urgent need for AI tools designed with accessibility at their core to ensure equitable participation in evolving coding practices.

\subsubsection{Design Recommendations.} Based on our findings, we propose a set of design recommendations aimed at improving interaction continuity, personalization, and information management in AI-assisted coding environments. These recommendations build on established principles such as \textit{Design for Generative Variability} by Weisz et al.~\cite{weisz2024design}, which emphasizes the importance of visualizing the user’s journey to accommodate diverse needs. From an Activity Theory perspective, our recommendations seek to realign the AI coding assistant (\textit{tool}) with developers who are visually impaired (\textit{subject}) to help them successfully complete coding tasks (\textit{object}).

First, we recommend implementing a \textbf{context history log}~\cite{byun2004utilizing} to mitigate workflow disruptions caused by AI-driven context switching. This feature would provide a sequential record of user interactions with the AI, allowing developers to review and revisit previous steps in their coding process. For visually impaired users, a structured history log could be especially valuable for maintaining workflow continuity and recovering from unexpected focus shifts.

Additionally, this log could enhance the system’s adaptability by learning from user behavior. For example, if a developer frequently accesses the inline chat for code explanations, the AI could prioritize and streamline access to this feature, reducing cognitive load and improving efficiency. By addressing workflow \textit{contradictions}, a context history log would ensure that developers maintain control over their interactions with the AI, making AI-assisted coding environments more predictable and user-friendly.

Second, we propose integrating an \textbf{interaction organizer} to help mitigate the challenges caused by the sequential navigation constraints of screen readers. Unlike sighted developers, who can visually scan and filter information quickly, developers who are visually impaired must process content in a linear fashion, which can make it difficult to efficiently navigate AI-generated suggestions \cite{borodin2010more}. The interaction organizer would allow users to structure AI-generated suggestions more effectively by enabling them to:

\begin{itemize}
\item Group related AI suggestions into folders or categories.
\item Assign meaningful labels and keywords for quick retrieval.
\item Attach personal notes or comments to AI-generated code. \end{itemize}

By incorporating these capabilities, the interaction organizer would help developers streamline their workflow, making it easier to manage and retrieve AI-generated outputs while reducing cognitive strain. From an Activity Theory perspective, this feature helps resolve \textit{contradictions} between the dynamic and non-linear nature of AI-generated suggestions and the sequential workflows that visually impaired developers rely on. By providing structured ways to categorize and personalize AI outputs, the interaction organizer improves the tool's role as a mediator in the coding process, making AI-assisted development more efficient and accessible.

\subsection{The Need for "AI Timeouts" in\\ AI-Assisted Coding}
Our study uncovered a complex relationship between increased productivity and cognitive overload in AI-assisted coding environments for developers who are visually impaired. While participants reported productivity gains—such as AI proactively anticipating their coding needs—these benefits also introduced a challenge: the need for \textit{AI timeouts}. These timeouts represent intentional breaks from AI interventions, allowing developers to manage information overload and maintain focus during their workflow.

The continuous stream of AI-generated suggestions, particularly inline \textit{ghost text}, seemed to have contributed to cognitive overload among participants. This overload created a need for \textit{AI timeouts}—temporary pauses in AI interventions to allow for uninterrupted focus and reduce mental strain.

Activity Theory provides a useful framework for understanding this challenge, as it reveals a \textit{contradiction} between the AI assistant’s intended role of enhancing productivity and its unintended effect of disrupting focus. This finding aligns with prior research on information overload in visual interfaces \cite{giraud2018web,ahmed2012read}, which shows that excessive information can overwhelm users, particularly those who are visually impaired. Our study extends this research by demonstrating how AI-generated content can intensify cognitive overload in AI-assisted coding environments. This underscores the importance of designing AI interactions that are more controlled and customizable, ensuring that developers who are visually impaired can maintain focus and effectively manage their workflows.

The need for \textit{AI timeouts} varied among participants, highlighting the complexity of designing AI coding assistants for developers who are visually impaired. While some participants valued proactive AI assistance, others wanted greater control over when and how AI intervened in their coding process. This variation underscores the importance of flexible, personalized customization rather than a one-size-fits-all approach, which is often emphasized in existing research on accessible development tools \cite{baker2015structjumper,potluri2018codetalk,potluri2022codewalk,armaly2018audiohighlight}.

Building on Gajos et al.'s \textit{adaptive interfaces} concept \cite{gajos2008improving}, our findings suggest that AI-assisted coding environments could benefit from adaptive features. From an Activity Theory perspective, these adaptive interfaces help resolve \textit{contradictions} by aligning AI mediation with user goals and preferences. Specifically, AI coding assistants could learn individual preferences for AI intervention, dynamically adjusting the frequency and type of AI suggestions based on task complexity. Additionally, providing easily accessible controls for enabling and disabling AI assistance would grant developers greater autonomy over their workflow. This approach ensures that AI remains a supportive tool rather than a source of cognitive overload.

\subsubsection{Addressing the Research Gap} Our research introduces \textit{AI timeouts} as a strategy to mitigate cognitive overload for developers who are visually impaired using AI coding assistants. Prior work highlights AI's ability to enhance productivity by automating repetitive tasks and offering anticipatory suggestions \cite{noy2023experimental,yu2024impact,al2024enhancing}. However, our findings reveal new challenges, such as cognitive strain from continuous AI-generated inputs like inline \textit{ghost text}. Building on studies of information overload for visually impaired users \cite{giraud2018web,ahmed2012read,peng2023impact,russo2024navigating}, we extend this discussion to dynamic AI-driven environments, where proactive AI mediation can disrupt cognitive coherence \cite{peng2023impact,russo2024navigating}. Using Activity Theory, we frame this tension as a \textit{contradiction} between AI assistance and focus disruption, proposing AI timeouts as a novel accessibility strategy to balance intervention and cognitive load.

\subsubsection{Design Recommendations.} To improve the accessibility and usability of AI coding assistants for developers who are visually impaired, we recommend design strategies aligned with the \textit{Design for Mental Models} principle by Weisz et al.~\cite{weisz2024design}. This principle emphasizes the importance of aligning AI behavior with the user’s expectations to balance supportive AI intervention with user autonomy. One of our recommendation is to allow developers to \textbf{customize the AI’s behavior} to match their mental model. For instance, tools could offer adjustable modes such as "Low Detail" and "High Detail," enabling users to control the level of information the AI provides based on task complexity or cognitive load. This flexibility ensures that AI assistance aligns with individual problem-solving preferences, reducing unnecessary cognitive strain. Customization can also resolve \textit{contradictions} between the user’s mental model and the often unpredictable nature of AI-generated suggestions, which may sometimes provide excessive or distracting information.

Another essential feature that we propose is the implementation of \textbf{AI timeouts}, which serve two key purposes:

\begin{itemize}
\item \textbf{Preserving Autonomy}: Timeouts allow developers to temporarily pause AI interventions, giving them space to reflect and solve problems independently. This can prevent over-reliance on AI and help maintain control over the coding process. From an Activity Theory perspective, these timeouts address \textit{contradictions} by giving users greater control over the tool, ensuring AI assistance aligns with their goals and cognitive needs.
\item \textbf{Enhancing Learning}: By pausing AI interventions, developers have the opportunity to engage with their own problem-solving strategies. This reinforces their mental model, strengthens coding skills, and promotes deeper understanding. This aligns with Activity Theory’s emphasis on fostering user agency, allowing developers to actively shape their interactions with Copilot.
\end{itemize}

For developers who are visually impaired, maintaining a clear and consistent mental model of the coding environment can be essential for effective work. Customizable AI behavior and strategic pauses could help ensure interactions remain predictable and aligned with user expectations, fostering a more intuitive and personalized experience. By supporting mental model alignment, these design features bridge the gap between user cognition and AI functionality, ultimately improving accessibility and productivity.

\subsection{Limitations and Future Work.} Our study provides valuable insights into the experiences of developers who are visually impaired using AI coding assistants. But it also has limitations. We interviewed a small group of 10 male participants, which restricts the diversity and generalizability of our findings. Studying emerging technologies like GitHub Copilot and recruiting a niche population, such as developers who are visually impaired, often results in a limited and homogeneous sample. However, despite participants' varying levels of experience with AI coding assistants, their shared challenges and opportunities provide foundational insights into how these tools impact accessibility.

Additionally, our study focused exclusively on GitHub Copilot, meaning we did not explore other AI coding assistants that may have different features and accessibility considerations. Future research should examine a broader range of AI coding tools to gain a more comprehensive understanding of accessibility challenges across different platforms. Another limitation is that participants had limited prior exposure to GitHub Copilot, meaning our findings may not fully capture the long-term benefits and challenges of using AI coding assistants. Longitudinal studies are needed to investigate how developers who are visually impaired adapt to AI-assisted environments over time, as well as any lasting accessibility barriers or professional impacts that may emerge.


\section{Conclusion}
We examined the experiences of 10 developers who are visually impaired as they interacted with an AI coding assistant. Our study highlights the dual impact of AI coding assistants on developers who are visually impaired. While these tools enhance control and provide valuable support, they also introduce new accessibility challenges and cognitive demands. Participants' experiences underscore the need to rethink accessibility in AI-driven coding environments. As AI continues to reshape software development, the design decisions made today will determine the inclusivity of the tech industry for years to come. Our findings call on researchers, designers, and industry leaders to make accessibility a priority in the development of next-generation AI tools, ensuring they empower all developers and foster equal opportunities in the field.

{\bf ACKNOWLEDGMENTS.} Special thanks to all the anonymous reviewers who helped us to strengthen the paper as well as the software developers who participated in the study. This work was partially supported by NSF grants 2339443 and  2403252.
\bibliographystyle{ACM-Reference-Format}
\bibliography{sample-base}

\end{document}